\documentclass[11pt]{article}

\usepackage{amssymb}
\usepackage{amsmath}
\usepackage{amsthm}
\usepackage[ruled,vlined,linesnumbered]{algorithm2e}

\usepackage{comment}
\usepackage{enumitem}
\usepackage{xcolor}
\usepackage{listings}
\usepackage{xcolor}
\usepackage{xspace}
\excludecomment{hidden}

\usepackage{mdframed}
\newmdtheoremenv[
    backgroundcolor=gray!5,
    linewidth=1pt,
    innerleftmargin=10pt,
    innerrightmargin=10pt
]{example}{Example}[section]
\newcommand{\ours}{\textsc{Self-Redraft}\xspace}

\usepackage[preprint]{acl}

\usepackage{times}
\usepackage{latexsym}
\usepackage{multirow}

\usepackage[T1]{fontenc}

\usepackage[utf8]{inputenc}

\usepackage{microtype}

\usepackage{inconsolata}

\usepackage{graphicx}
\usepackage{booktabs}

\usepackage{adjustbox}   

\usepackage{tcolorbox}
\tcbuselibrary{skins, breakable}
\newtcolorbox[auto counter, number within=section]{promptbox}[2][]{%
    colframe=blue!75!black,
    colback=blue!10,
    coltitle=white,
    fonttitle=\small\bfseries,
    title=Prompt Template~\thetcbcounter: #2,
    breakable, 
    enhanced,
    fontupper=\ttfamily,
    #1 
}

\title{\ours: Eliciting Intrinsic Exploration-Exploitation Balance in Test-Time Scaling for Code Generation}

\author{Yixiang Chen\thanks{Equal Contribution.}, Tianshi Zheng\footnotemark[1], Shijue Huang, Zhitao He,  Yi R. (May) Fung \\
Department of Computer Science and Engineering, HKUST\\
\texttt{ychenla@connect.ust.hk, yrfung@cse.ust.hk}}

\NewDocumentCommand{\tianshi}
{ mO{} }{\textcolor{orange}{\textsuperscript{\textit{tianshi}}{{\small[#1]}}}}

\begin{document}
\maketitle

\begin{abstract}

Test-time scaling without interpreter feedback is essential for real-world code generation scenarios where test cases are not readily available. While existing paradigms often rely on either greedy exploitation (i.e., iterative refinement) or stochastic exploration (i.e., relying on sample-based voting or reranking mechanisms), the balance between these two dimensions remains underexplored. To investigate the LLM's intrinsic ability to balance exploitation and exploration, we introduce \ours, a framework built upon \underline{Self-Re}fine that encourages the model to propose new \underline{draft}s for solutions that are fundamentally flawed.

Our results show that \ours consistently achieves better performance than Self-Refine when converged under the same maximum number of iterations.
Still, we observe that significant room for improvement remains, largely due to two core aspects of current self-redraft capabilities: constrained capacity for generating instructive feedback and fragile discriminative judgment. We also find that balancing strategies vary notably across different LLMs, reflecting distinct, model-specific behaviors. Overall, our study establishes a baseline for intrinsic exploration-exploitation balancing in test-time scaling and identifies feedback and discrimination as key areas with potential for future advances.
\footnote{Our code will be released publicly upon publication.}

\end{abstract}
\begin{figure}[t]
    \centering
    \includegraphics[width=\linewidth]{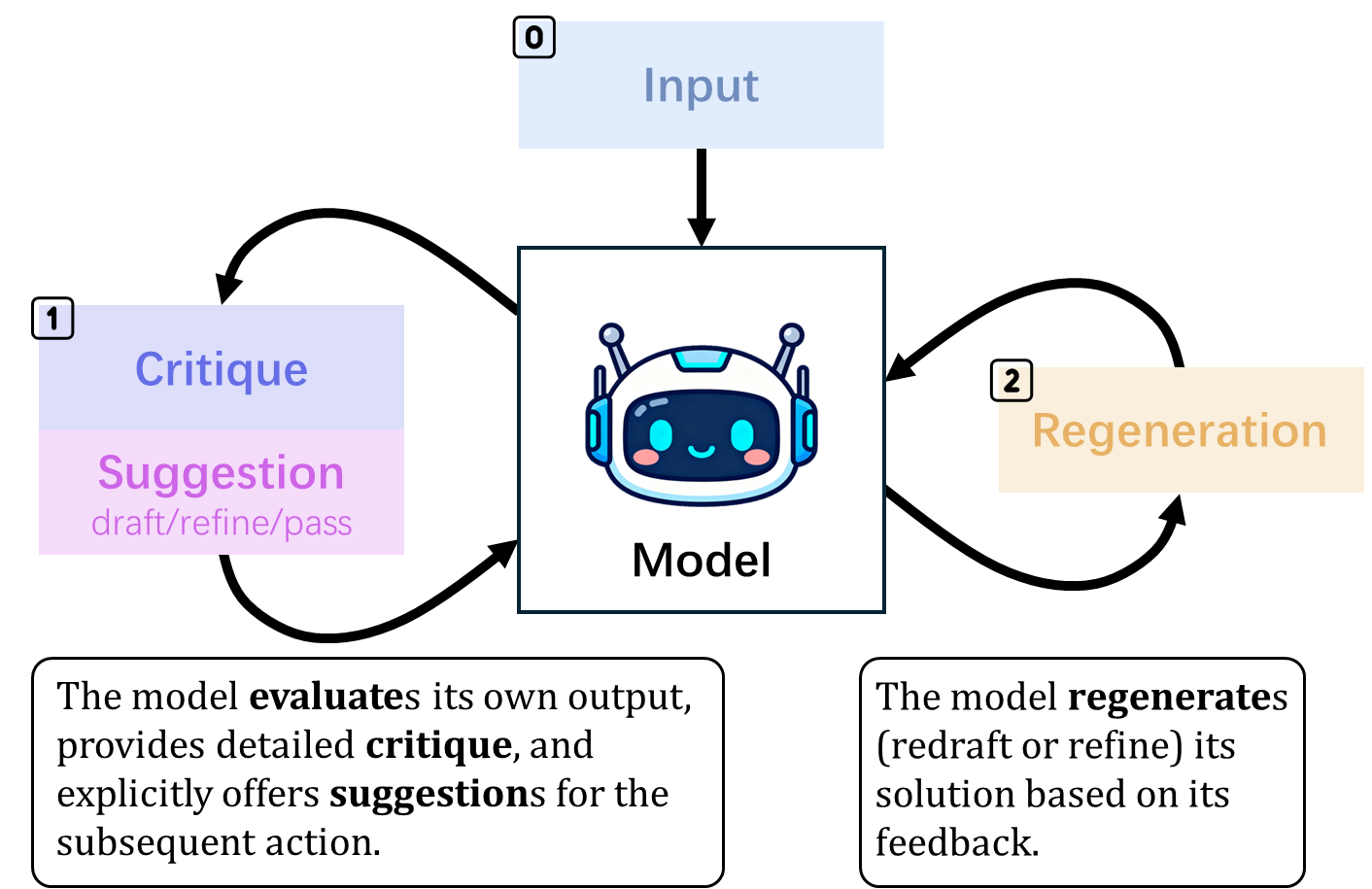}
\caption{ Our proposed \ours framework.} 
    \label{fig:self-redraft}
    \vspace{-0.3cm}
\end{figure}
\begin{figure*}[!htbp]
    \centering
    \includegraphics[width=\textwidth]{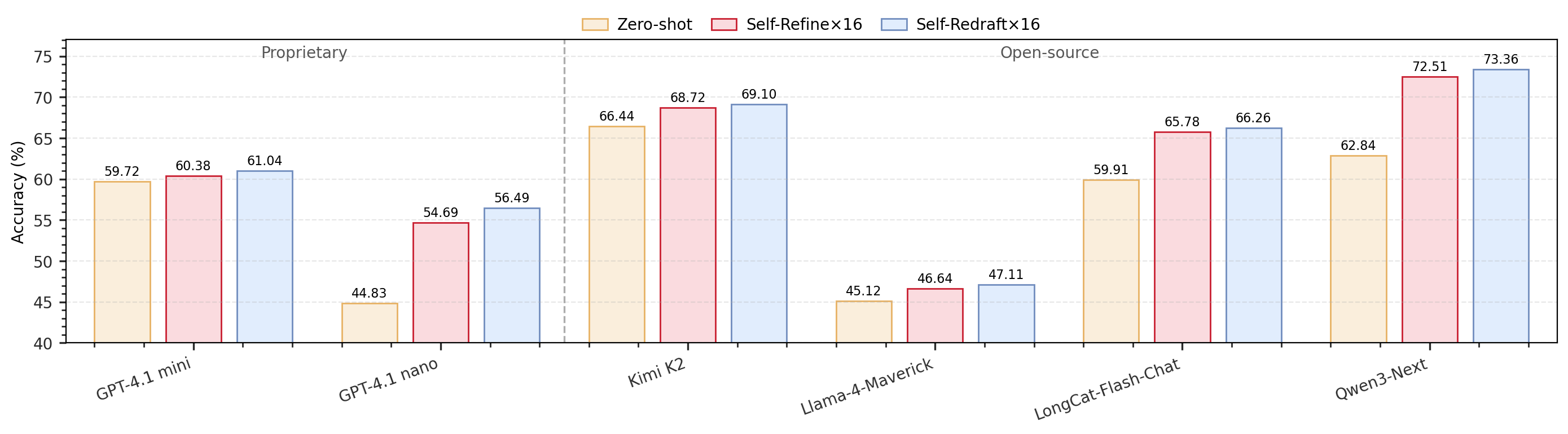}
    \vspace{-2.1em}
    \caption{Detailed benchmark performance of LLMs on LiveCodeBench, evaluated with 16 iterations of Self-Refine and 16 iterations of \ours.}
    \label{fig:main_result}
    \vspace{-0.3cm}
    
\end{figure*}

\section{Introduction}
Increasing test-time compute has emerged as a powerful means of boosting large language model (LLM) performance in code generation \citep{snell2024scaling, OpenAI2024learning, guo2025deepseek, brown2024large}, with prevailing test-time scaling approaches broadly categorized into execution-based \citep{li2025s, jin2025reveal, shinn2023reflexion, chen2023teaching, jiang2025aide} and execution-free \citep{chen2025sets, xie2025teaching} frameworks. 

Existing test-time scaling paradigms can be framed as a tree search, inherently balancing greedy exploitation with stochastic exploration \citep{tang2024code}. This \textbf{exploration-exploitation tradeoff}—navigating between refining current solutions (exploitation) and proposing novel ones (exploration)—is central to overcoming local optima and enhancing solution diversity.

While significant research has focused on balancing this tradeoff within \textbf{execution-based} frameworks \citep{li2025s, jin2025reveal, shinn2023reflexion, chen2023teaching, jiang2025aide, liu2025ml}, \textbf{execution-free} methods \citep{xie2025teaching, chen2023teaching} have predominantly relied on exploitation, leaving the exploration aspect of test-time scaling largely unaddressed. This represents a critical gap, as real-world scenarios often lack the readily available test cases required for execution-based methods \citep{ouedraogo2025enriching,huang2025benchmarking}.

To address this gap, we introduce \ours, a framework built upon the purely exploitative \textsc{Self-Refine} process \citep{madaan2023self}. \ours introduces an explicit exploratory choice by modifying the feedback stage. Instead of only generating refinement instructions, the model is prompted to first diagnose its solution. If it identifies a fundamental flaw, it is encouraged to \textbf{redraft} an entirely new solution, effectively choosing exploration over exploitation.

Extensive experiments on LiveCodeBench \cite{jain2024livecodebench} demonstrate that \ours \textbf{consistently outperforms} the purely exploitative Self-Refine within the same iteration budget. We also observe that a notable performance gap remains compared to the pass@k upper bound, which reflects the potential of pure exploration. This suggests that the primary challenge lies in the LLM's intrinsic capacity for self-guided exploration, limiting its ability to fully leverage the substantial untapped potential in this area.
 
To better understand the factors behind this gap, we examine two core bottlenecks in LLMs' intrinsic capabilities:
(1) \textbf{Limited Feedback Generation}: The models struggle to produce sufficiently critical feedback. Consequently, they often fail to recognize when a solution is fundamentally flawed and requires a complete redraft rather than an incremental refinement.
(2) \textbf{Fragile Discriminative Judgment}: The models' ability to distinguish correct code from incorrect code is unreliable. This fragility causes them to erroneously ``refine'' correct solutions into incorrect ones or to accept flawed solutions.
We further observe that these behaviors differ considerably across LLMs, indicating that the balancing of exploration and exploitation is not yet a generalizable capability in current models, but rather an emergent and model-specific characteristic.


These findings establish a baseline for intrinsic exploration-exploitation balancing, pointing to critical directions for future research: improving feedback generation, enhancing discriminative judgment, and designing model-adaptive strategies.
\section{Methodology}
\vspace{-0.6em}
Our work focuses on Execution-Free Test-Time Scaling for Code Generation. Typically, the success of test-time scaling hinges on maximizing the utilization of two key search modes: exploitation and exploration. In this work, we leverage \ours as tools to examine the model's intrinsic ability to balance exploitation and exploration.

\subsection{Test-time Scaling with \ours}
\vspace{-0.3em}
Similar to Self-Refine, \ours, as illustrated in Figure \ref{fig:self-redraft}, iterates between feedback and regeneration until a stopping condition is met. The entire framework  consists of three main steps: Step 0: Given a programming task $x$ and a generation prompt $p_{\text{gen}}$, the model first produces an initial solution $y_0 \sim \pi(\cdot \mid p_{\text{gen}}, x)$.  
Step 1: The model evaluates its solution $y_i$ and, using the feedback prompt $p_{\text{fb}}$, generates feedback $c_i \sim \pi(\cdot \mid p_{\text{fb}}, x, y_i)$ that includes critique and an explicit suggestion for the next action (i.e., draft, refine, or pass).  
Step 2: Based on prior feedback and solutions, the model regenerates (either redrafts or refines) a new solution $y_{i+1} \sim \pi(\cdot \mid p_{\text{regen}}, x, y_i, c_i, \ldots, y_0, c_0)$ using the regeneration prompt $p_{\text{regen}}$.  
Steps 1 and 2 iterate until a predefined stopping condition is satisfied. The core difference with Self-Refine, however, lies in the approach to search modes: \ours explicitly encourages the model to produce a fresh draft for solutions identified as methodologically flawed. Details of the algorithm and prompts are provided in the Appendix \ref{sec:self-redraft}. This design enables the model to simultaneously leverage both exploitation and exploration in test-time scaling.

\subsection{Dataset and Models}
\vspace{-0.3em}
\paragraph{Dataset. } We conduct our experiments on the latest version of LiveCodeBench \citep{jain2024livecodebench}, which offers a holistic and contamination-free evaluation of the coding capabilities of LMs. It comprises 1,055 programming problems, categorized into three difficulty levels: easy, medium, and hard.

\paragraph{Models. } We evaluate  6 open-source and  proprietary LLMs with various parameter sizes. Configurations are detailed in Appendix \ref{sec:model-details}.

\begin{table*}[!htbp]
\centering
\small
\setlength{\tabcolsep}{8pt}
\renewcommand{\arraystretch}{1.1}
\begin{tabular}{lccccc}
\toprule
\textbf{Model} & \multicolumn{2}{c}{\textbf{Improvement Rate} $r_{\text{imp}}$ (\%)} & & \multicolumn{2}{c}{\textbf{Regression Rate} $r_{\text{reg}}$ (\%)} \\
\cmidrule(lr){2-3} \cmidrule(lr){5-6}
 & Self-Refine & \ours & & Self-Refine & \ours \\
\midrule
GPT-4.1 mini       & 3.29  & 5.18 {\scriptsize (\textcolor{red}{+1.89})}  & & 1.11 & 1.27 {\scriptsize (\textcolor{red}{+0.16})} \\
GPT-4.1 nano       & 19.52 & 23.02 {\scriptsize (\textcolor{red}{+3.50})} & & 1.70 & 2.33 {\scriptsize (\textcolor{red}{+0.63})} \\
Kimi K2            & 9.89  & 12.99 {\scriptsize (\textcolor{red}{+3.10})} & & 1.57 & 2.57 {\scriptsize (\textcolor{red}{+1.00})} \\
Llama-4-Maverick   & 4.15  & 6.74 {\scriptsize (\textcolor{red}{+2.59})}  & & 1.68 & 3.78 {\scriptsize (\textcolor{red}{+2.10})} \\
LongCat-Flash-Chat & 18.68 & 20.33 {\scriptsize (\textcolor{red}{+1.65})} & & 2.69 & 3.01 {\scriptsize (\textcolor{red}{+0.32})} \\
Qwen3-Next         & 26.53 & 29.34 {\scriptsize (\textcolor{red}{+2.81})} & & 0.30 & 0.60 {\scriptsize (\textcolor{red}{+0.30})} \\
\bottomrule
\end{tabular}
\caption{Improvement ($r_{\text{imp}}$) and regression ($r_{\text{reg}}$) rates  of \ours and Self-Refine over 16 iterations. }
\label{tab:imp_reg}
\vspace{-0.3cm}
\end{table*}


\section{Experiments and Analysis}
\subsection{Main Results}
\label{subsec:main_results}
We evaluate each model on LiveCodeBench using Self-Refine \citep{madaan2023self} and \ours with iterations ranging from 1 to 16 (Self-Refine x16 and \ours x16), where the iterations of both Self-Refine and \ours are based on the same set of initial solutions.
The main experimental results are illustrated in Figure \ref{fig:main_result} (full results in Appendix \ref{sec:full-results}).
As results stabilize by 16 iterations, \ours achieves a modest yet consistent average absolute gain of 0.615\% over Self-Refine. 
Details of Self-Refine are provided in Appendix \ref{sec:self-refine}.

\subsection{Unexploited Potential in Exploration}
\begin{figure}[hbp]
    \centering
    \vspace{-1em}
    \includegraphics[width=\linewidth]{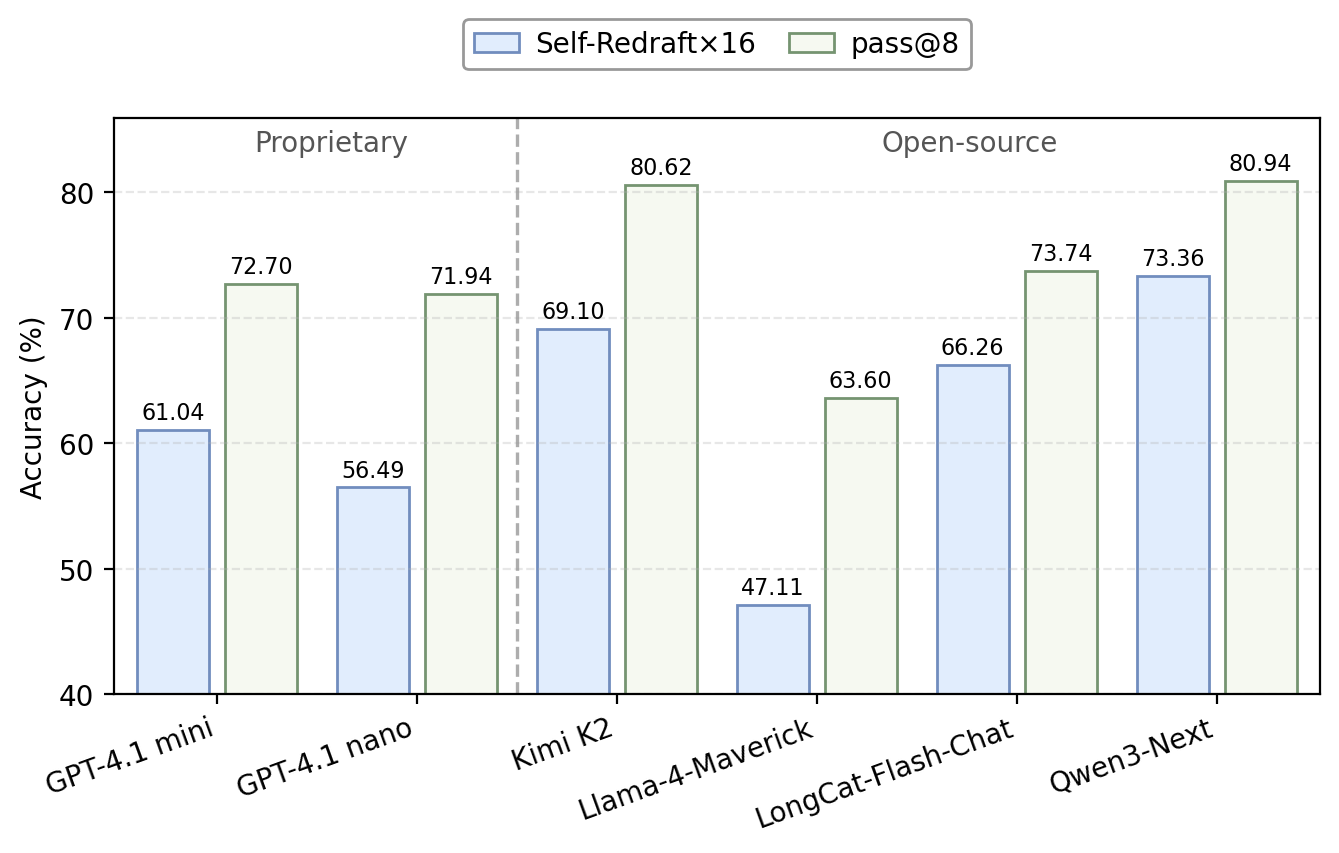}
    \vspace{-2.1em}
    \caption{Comparison of \ours×16 and pass@8 accuracies on LiveCodeBench.}
    \vspace{-0.5em}
    \label{fig:passat8}
\end{figure}

To contextualize this gain and assess the room for improvement, we compare the performance of \ours x16 against the pass@8 upper bound using 16 samples, which reflects the potential of pure exploration, as presented in Figure \ref{fig:passat8}.
The strength of pass@8 suggests that exploration alone is highly promising for test-time scaling: for a substantial subset of problems, eight initial (unrefined) samples suffice to contain a correct solution. Nevertheless, \ours, which aims to balance exploration and exploitation in an execution-free setting, fails to exploit this advantage effectively even when allowed to generate up to 17 solutions within 16 iterations. These results motivate a systematic investigation into execution-free approaches that more effectively allocate sampling budgets, diversify candidates, and select among them to close the gap.
Details of the pass@k metric are provided in Appendix \ref{sec:self-refine}.
\subsection{Insufficient Model Critique for Methodological Correction
}
\label{subsec:critique}
Prior work has shown that large language models (LLMs) often struggle to produce feedback that is both informative and actionable \citep{zheng2024makes, xie2025teaching}. Building on these observations, we investigate whether, within \ours, models can reliably distinguish cases that warrant incremental refinement of the current solution from those that require drafting a new solution from scratch, and whether their feedback appropriately recommends the corresponding intervention. To this end, we conduct a blinded evaluation of model-generated critiques, assessing their understanding in prescribing refine versus redraft with actionable guidance.

\paragraph{Blinded Evaluation. }
Following \citet{xie2025teaching} who characterized critique ability through Markov chain transition dynamics \citep{meyn2012markov} analyzing solution correctness before and after refinement, we similarly leverage methodological changes between original and regenerated solutions to examine whether models provide effective feedback recommendations. 

We sample solution pairs from trajectories collected in Section \ref{subsec:main_results} and engage various auxiliary models in a blinded evaluation. These models are presented solely with solution pairs (before and after regeneration) and asked to annotate whether methodological changes occurred. We then compare these annotations against the actual next-step actions recommended in the original feedback \citep{tan2024large}. To ensure balanced evaluation, we maintain equal representation of ``draft'' and ``refine'' labels within each sampled group. Comprehensive experimental details and results are provided in Appendix \ref{sec:blinded-evaluation}.

\begin{figure}[htbp]
    \centering
    \includegraphics[width=\linewidth]{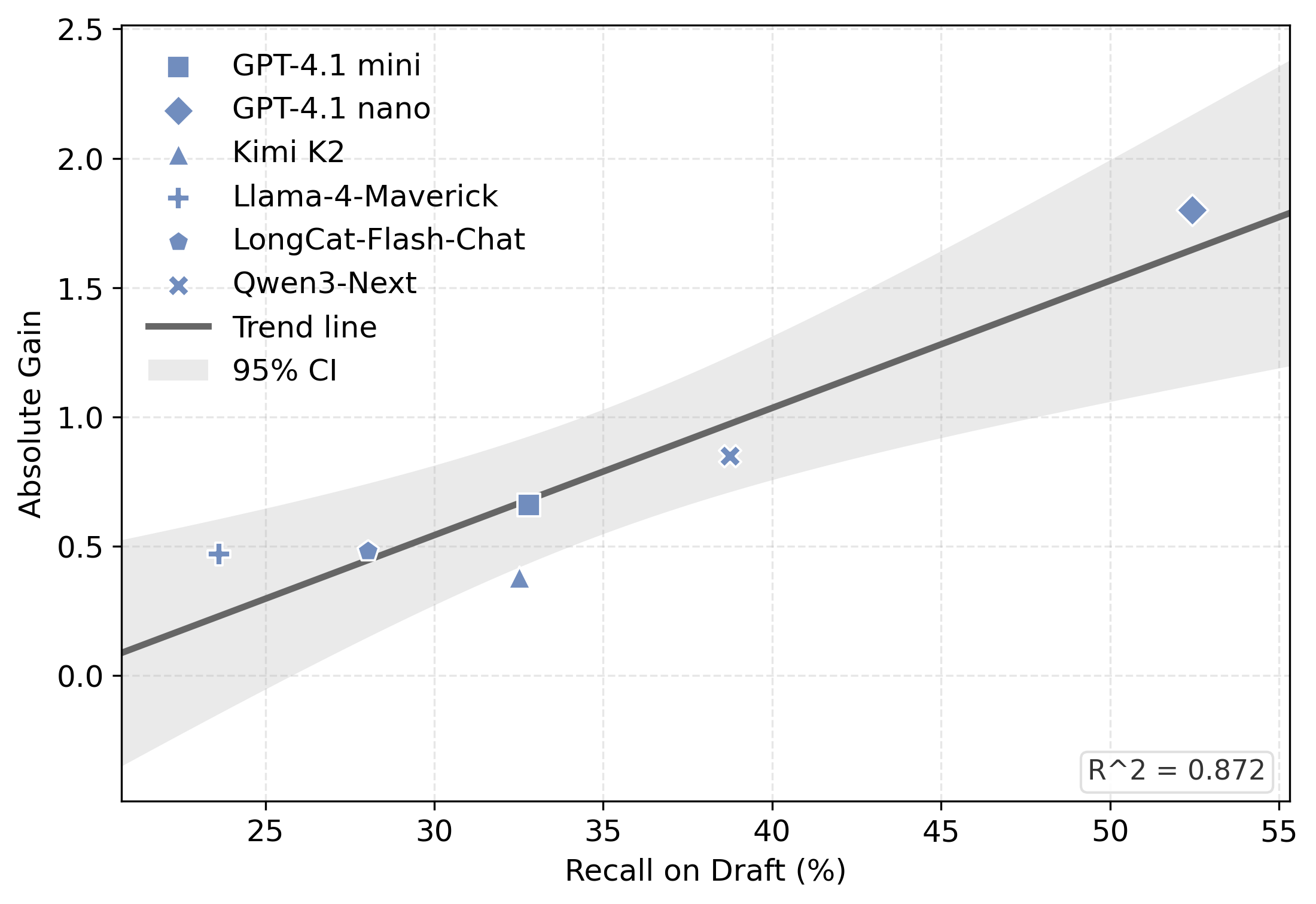}
    \vspace{-2.1em}
    \caption{\small Recall on Draft versus absolute improvement of \ours x16 over Self-Refine x16.}
    \label{fig:scatter_plot}
    \vspace{-0.3cm}
\end{figure}

We employ ``Recall on Draft'' to measure how often auxiliary evaluators correctly identify feedback that recommended a substantive methodological change (“draft”). Average recall values per model are shown in Figure \ref{fig:avg_recall}. Notably, Recall on Draft exhibits a positive correlation with the absolute improvement of \ours over Self-Refine (Figure \ref{fig:scatter_plot}). Furthermore, the ranking of models by recall remains largely consistent across different evaluators (Figure \ref{fig:ranking}), indicating a shared understanding of methodological shifts. This consistency supports the inference that most models fail to provide actionable feedback for methodological correction, thereby limiting effective exploration.

\subsection{Fragile Code Discrimination Leading to Deleterious Redrafts}
Besides the identified limitations in model critique, we further hypothesize that the limited performance improvement may be attributed to the model's inherent difficulty in assessing code correctness. To verify this, we compare two key metrics across \ours and Self-Refine—over 16 iterations and across various generator models using the same set of initial solutions: the improvement rate (\(r_{\text{imp}}\), proportion of initially incorrect solutions corrected) and the regression rate (\(r_{\text{reg}}\), proportion of initially correct solutions corrupted). Results are presented in Table \ref{tab:imp_reg}. We find that while \ours exhibits a higher \(r_{\text{imp}}\) (i.e., corrects more initially incorrect solutions) than Self-Refine, it simultaneously demonstrates a notably higher \(r_{\text{reg}}\) (i.e., corrupts more initially correct solutions). This observation underscores that, constrained by the models’ limited discriminative capacity, redrafting solutions via alternative approaches—a core design of \ours—becomes a high-risk action prone to deleterious outcomes.

\subsection{Cross-Model Inconsistency}
Beyond the two key findings outlined above—insufficient model critique for methodological correction and fragile code discrimination leading to deleterious redrafts—we further observe substantial variation in exploration-exploitation balancing strategies across different generator models, as illustrated in Figure \ref{fig:butterfly}. Such divergence reflects inconsistent understandings among models regarding how to navigate the exploration-exploitation tradeoff: some models prioritize ``refine'' actions (leaning toward exploitation), while others favor ``redraft'' suggestions (tilting toward exploration), with no cohesive pattern across the cohort. This lack of consistency underscores that, as general-purpose LLMs are not specifically designed to manage this tradeoff, their intrinsic ability to maintain a robust balance between exploration and exploitation remains insufficient.


\section{Conclusion}
\vspace{-0.3em}
We introduced \textsc{Self-Redraft}, a framework that elicits an intrinsic balance between exploration and exploitation in execution-free code generation. Experiments demonstrate that \textsc{Self-Redraft} consistently outperforms the purely exploitative Self-Refine under a comparable iteration budget. However, our analysis reveals that the primary performance bottleneck is not the model's generative breadth, but its capacity for accurate self-diagnosis and strategic decision-making. This is evidenced by two key challenges: a limited ability to generate useful feedback for methodological correction and fragile discriminative judgment that leads to deleterious redrafts. We also find that balancing strategies vary significantly across different LLMs, indicating that this capability is model-specific rather than general. Overall, our work establishes a baseline for intrinsic exploration-exploitation balancing and identifies robust self-reflection as a critical frontier for advancing execution-free code generation.



\section*{Limitations}
While our study provides a systematic analysis of the challenges in balancing exploration and exploitation, it is subject to several limitations that also highlight avenues for future research:

\begin{itemize}
    \item \textbf{Execution-Free Paradigm:} Our work is intentionally restricted to an execution-free setting to study the intrinsic capabilities of LLMs. This focus, however, means our findings are not directly comparable to execution-based methods, and future work could investigate hybrid approaches that bridge this gap.

    \item \textbf{Benchmark Generalizability:} The experiments are conducted exclusively on LiveCodeBench. Although it is a comprehensive benchmark, the generalizability of our findings to other coding domains, programming languages, or problem types remains to be verified.

    \item \textbf{Reliance on Intrinsic Capabilities:} The \textsc{Self-Redraft} framework is designed to elicit the \textit{intrinsic} self-correction abilities of pre-trained models. Its performance is therefore constrained by these inherent capabilities. We do not explore training-driven improvements (e.g., fine-tuning models to be better critics) or alternative, non-intrinsic exploration strategies, which represent valuable directions for future investigation.
\end{itemize}

\section*{Ethics Statement}
Our work focuses on analyzing the intrinsic capabilities of existing LLMs for code generation. We used pre-trained models accessible via public APIs or releases and did not train new models, thereby limiting the direct computational and environmental footprint of our study. All experiments were conducted on LiveCodeBench, a publicly available and contamination-free benchmark, ensuring the integrity and reproducibility of our evaluation. We acknowledge the broader ethical concerns associated with code-generating LLMs, including the potential for generating insecure or biased code and issues related to intellectual property. Our research, however, is analytical in nature, aiming to understand the fundamental mechanisms of self-correction rather than deploying a generative system. The insights gained are intended to contribute to the development of more reliable and transparent models. All evaluations, including the blinded study, were performed using auxiliary models, with no human annotation involved.

\bibliography{custom}

\clearpage

\appendix

\section{\ours}
\label{sec:self-redraft}
\subsection{Algorithm}
See Algorithm \ref{alg:self-redraft}. In our setup, the stopping criteria for \ours are defined as follows: either the number of iterations reaches the preset maximum, or the model explicitly responds with ``pass'' during the feedback phase. Consequently, in most cases, the model does not reach the maximum number of iterations—this is due to early termination triggered by the model’s explicit ``pass'' response.
\subsection{Prompt}
For generation prompts, we use the default settings of LiveCodeBench. For feedback prompts, in both Self-Refine and \ours, we instruct models to leverage XML tags to generate structured outputs—this design facilitates information extraction, algorithmic decision-making within the framework, and subsequent experimental analyses.
Aligned with the Self-Refine algorithm, \ours exhibits distinct prompt designs for feedback and regeneration: for feedback prompts, only the previous solution is presented to the model; for regeneration prompts, by contrast, all prior solutions and corresponding feedback are incorporated in the form of a trajectory.

The detailed prompt instructions are provided below:

\begin{promptbox}[colback=black!5, colframe=white!60!black, title=Prompt Templates]{}
\textbf{\small{Feedback Prompt in Self-Refine}}
\scriptsize
\vspace{0.2cm}

You are an expert Python programmer. You will be given a 
question and a piece of code. Check if the code correctly 
solves the problem and passes all examples. Provide feedback ONLY (no extra content).
\vspace{0.2cm}

\begin{verbatim}
## Task:
<task description>

## Code:
<previous solution>

## Note:\end{verbatim}
\begin{enumerate}[leftmargin=*]
    \item Your feedback should consist of two parts: critique and
suggestion. In the critique, you should analyze the code 
and provide specific recommendations. In the suggestion, 
you should clarify the direction for the next steps.
    \item Your suggestion should be one of ``pass'' and ``refine''.
    \item If you think the code is correct, your suggestion should
be ``pass''.
    \item If you think the code should be refined (small improvements), offer guidance on the refinement and your suggestion should be ``refine''.
\end{enumerate}

\begin{verbatim}
## Format: 
<critique>
your detailed critique and analysis here
</critique>
<suggestion>
pass/refine
</suggestion>
\end{verbatim}

\hrule height 0.5pt
\vspace{0.3cm}
\textbf{\small{Feedback Prompt in \ours}}
\scriptsize
\vspace{0.2cm}

You are an expert Python programmer. You will be given a 
question and a piece of code. Check if the code correctly 
solves the problem and passes all examples. Provide feedback ONLY (no extra content).
\vspace{0.2cm}
\begin{verbatim}
## Task:
<task description>

## Code:
<previous solution>

## Note:\end{verbatim}
\begin{enumerate}[leftmargin=*]
    \item Your feedback should consist of two parts: critique and
suggestion. In the critique, you should analyze the code 
and provide specific recommendations. In the suggestion, 
you should clarify the direction for the next steps.
    \item Your suggestion should be one of ``pass", ``refine", and ``redraft".
    \item If you think the code is correct, your suggestion should
be ``pass''.
    \item If you think the code should be refined (small improvements), offer guidance on the refinement and your suggestion should be ``refine''.
    \item If the solution is fundamentally incorrect and needs a new approach, then encourage an alternative method to address the issue in the feedback, offer guidance on the new method and set your suggestion to ``redraft".
\end{enumerate}
\begin{verbatim}
## Format: 
<critique>
your detailed critique and analysis here
</critique>
<suggestion>
pass/refine/redraft
</suggestion>
\end{verbatim}

\end{promptbox}

\begin{promptbox}[colback=black!5, colframe=white!60!black, title=Prompt Templates]{}
\textbf{\small{Regeneration Prompt}}
\scriptsize
\vspace{0.2cm}

You are an expert Python programmer. Regenerate the code based on the feedback to solve the problem correctly. Follow the original formatting requirements.

\begin{verbatim}
## Task:
<task description>

## Iteration History:
<trajectory>

## Current Feedback:
<feedback>

## Regenerated Code: 
\end{verbatim}
\end{promptbox}

\begin{algorithm*}[!htbp]
\caption{\ours: Execution-free Test-time Scaling with Drafting}
\label{alg:self-redraft}
\DontPrintSemicolon
\SetKwInOut{Input}{Input}\SetKwInOut{Output}{Output}
\SetKw{KwReturn}{return}
\SetKw{KwBreak}{break}
\SetKw{KwTo}{to}
\SetKw{KwAnd}{and}

\Input{%
  Programming task $x$; generation prompt $p_{\text{gen}}$;\\
  feedback prompt $p_{\text{fb}}$; regeneration prompt $p_{\text{regen}}$;\\
  maximum iterations $T$.
}
\Output{Final solution $\hat{y}$.}

\textbf{Step 0 (Initialization):} Generate initial solution
$y_0 \sim \pi(\cdot \mid p_{\text{gen}}, x)$; set $i \leftarrow 0$.\;

\While{$i < T$}{
  \textbf{Step 1 (Feedback/Critique):}
  Generate critique and next-action suggestion
  $c_i \sim \pi(\cdot \mid p_{\text{fb}}, x, y_i)$,\\
  where $c_i$ includes (a) critique and (b) action
  $\in \{\textsc{draft}, \textsc{refine}, \textsc{pass}\}$.\;

  \If{action in $c_i$ is \textsc{pass}}{
     \KwReturn $\hat{y} \leftarrow y_i$\;
  }

  \textbf{Step 2 (Regeneration):}\;
  \uIf{action in $c_i$ is \textsc{draft}}{
     Generate a fresh solution:
     $y_{i+1} \sim \pi(\cdot \mid p_{\text{regen}}, x, y_i, c_i, \ldots, y_0, c_0)$.\;
  }
  \ElseIf{action in $c_i$ is \textsc{refine}}{
     Generate a refined solution:
     $y_{i+1} \sim \pi(\cdot \mid p_{\text{regen}}, x, y_i, c_i, \ldots, y_0, c_0)$.\;
  }

  $i \leftarrow i + 1$.\;
}

\KwReturn $\hat{y} \leftarrow y_i$ \tcp*[r]{Reached max iterations $T$}
\end{algorithm*}

\section{Preliminaries}
\begin{itemize}[leftmargin=*] 
    \item \textbf{Self-Refine} \citep{madaan2023self} is an iterative self-refinement algorithm that alternates between two generative steps---feedback and refine. Given an input sequence, Self-Refine generates an initial output, provides feedback on the output, and refines the output according to the feedback. Self-Refine iterates between feedback and refinement until a desired condition is met. In line with this, \ours is also proposed to be grounded in the model’s own feedback-refinement iterative logic—mirroring the core alternating structure of feedback and refinement steps that defines Self-Refine.
    \item \textbf{Pass@k} \citep{kulal2019spoc, chen2021evaluating} is a metric for functional correctness evaluation. First, $n$ samples are generated per task. Then, we count the number of correct samples $c$ and calculate the unbiased estimator:
    \begin{equation}
    \label{eq:passatk}
    \text{pass@k}=\underset{\text{Problem}}{\mathbb{E}}\left[1-\frac{\binom{n-c}{k}}{\binom{n}{k}}\right]
    \end{equation}
    In this paper, we use $n=16$ and $k=8$.
\end{itemize}
\label{sec:self-refine}
\section{Model Details}
In our experiments, we evaluated 6 modern LLMs with various parameter sizes. For all evaluations on LiveCodeBench, we set the model’s temperature to 0.2, top-$p$ to 0.95, and both frequency penalty and presence penalty to 0—all of which align with LiveCodeBench’s default parameters.
\label{sec:model-details}
\begin{itemize}[leftmargin=*]
    \item \textbf{GPT-4.1 mini} \citep{OpenAI2025introducing} is a mid-sized model delivering competitive performance at substantially lower latency and cost.
    \item \textbf{GPT-4.1 nano} \citep{OpenAI2024learning}  is the fastest and cheapest model in the GPT-4.1 series for tasks that demand low latency.
    \item \textbf{Kimi K2} \citep{kimiteam2025kimik2openagentic} is a state-of-the-art mixture-of-experts (MoE) language model with 32 billion activated parameters and 1 trillion total parameters. 
    \item \textbf{Llama 4 Maverick} \citep{Meta2025llama} is a mixture-of-experts (MoE) language model with 17 billion active parameter and 128 experts.
    \item \textbf{Longcat-Flash-Chat} \citep{meituan2025longcatflashtechnicalreport} is a non-thinking foundation mixture-of-experts (MoE) model that delivers highly competitive performance with exceptional strengths in agentic tasks.
    \item \textbf{Qwen3-Next-80B-A3B-Instruct} \citep{qwen3technicalreport} is an instruction-tuned chat model in the Qwen3-Next series optimized for fast, stable responses without ``thinking'' traces.
\end{itemize}
The above are the generator models used in our evaluation. In the blinded evaluation described in Section \ref{subsec:critique}, we utilized auxiliary models for annotation. The details of these auxiliary models are as follows:

\begin{itemize}[leftmargin=*]
    \item \textbf{GPT-5 mini} \citep{OpenAI2025introducing2}  is a compact version of GPT-5, designed to handle lighter-weight reasoning tasks.
    \item \textbf{GLM-4.6} \citep{z2025glm}  is a foundation mixture-of-experts (MoE) model designed for intelligent agents.
    \item \textbf{Grok 4 Fast} \citep{xAI2025grok} is xAI's latest multimodal model with state-of-the-art cost-efficiency and a 2M token context window. 
\end{itemize}

\section{Experiment Detail}
\subsection{Main Experiment}
For each generator model, we first generate an initial set of solutions. As preliminary experiments revealed that iterative performance is significantly influenced by the correctness of initial solutions, we conduct both Self-Refine and \ours using the same set of initial solutions to ensure fair comparison. In both settings, models are allowed up to 16 iterations, with early termination if the model suggests ``pass'', as outlined in Algorithm \ref{alg:self-redraft}. During the experiment, we collect full trajectories for each problem instance, including feedback and regenerated solutions at each iteration for subsequent analysis. All solutions from every iteration are evaluated for functional correctness (detailed results in Appendix \ref{sec:full-results}); for early-stopped trajectories, the final regenerated solution is used for assessment. The pass@8 metric is computed by generating 16 independent samples per task and evaluating them according to Equation \ref{eq:passatk}.
\subsection{Blinded Evaluation}
To investigate whether models effectively provide self-guidance for methodological correction, we conducted a blinded evaluation using several auxiliary models (see model details in Appendix \ref{sec:model-details}). For each generator model, we sampled (solution, feedback, regenerated solution) tuples from trajectories collected in our main experiments. The auxiliary models were presented with only the solution pairs (original and regenerated) and asked to classify whether the original feedback recommended ``refine'' or ``redraft''. To ensure balanced evaluation, we maintained equal representation of ``redraft'' and ``refine'' labels in our sampling, with a maximum of 1,000 samples per generator model. We evaluated classification performance using accuracy and Recall on Draft—the proportion of actual ``redraft'' instances that were correctly identified by the auxiliary models. The prompt used for the auxiliary models is as follows:
\vspace{0.2cm}
\begin{promptbox}[colback=black!5, colframe=white!60!black, title=Prompt Templates]{}
\textbf{\small{Annotation Prompt}}
\scriptsize
\vspace{0.2cm}

You are given an original solution and an alternative suggestion for the same task.

\vspace{0.2cm}
Classify the suggestion as either a minimal refinement of the original (small edits that preserve the solution's overall structure and approach) or a full redraft (substantial rewrite with a different structure/approach). 

\vspace{0.2cm}
Output only one XML tag exactly as 
\begin{verbatim}
<suggestion>refine</suggestion>
\end{verbatim}
or 
\begin{verbatim}
<suggestion>redraft</suggestion>.
\end{verbatim}

\begin{verbatim}
[Original Solution]
<original solution>

[Suggested Alternative]
<regenerated solution>
\end{verbatim}
Answer strictly with the XML tag only.
\end{promptbox}

\label{sec:blinded-evaluation}
\begin{figure}[htbp]
    \centering
    \includegraphics[width=\linewidth]{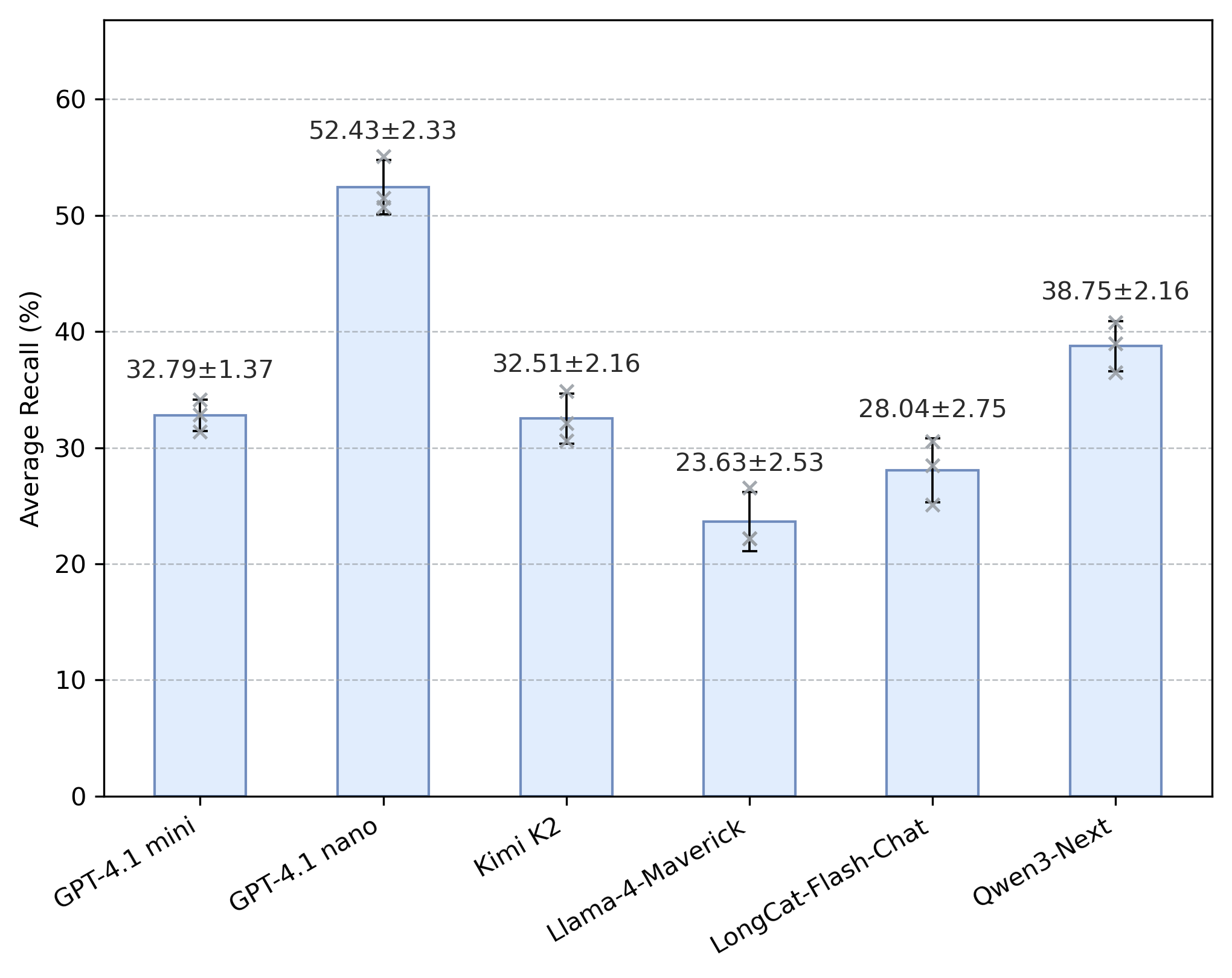}
    \caption{Recall on Draft as annotated by three auxiliary models: GPT-5 mini, GLM-4.6 and Grok 4 Fast. }
    \label{fig:avg_recall}
    \vspace{-0.3cm}
\end{figure}
\begin{figure}[htbp]
    \centering
    \includegraphics[width=\linewidth]{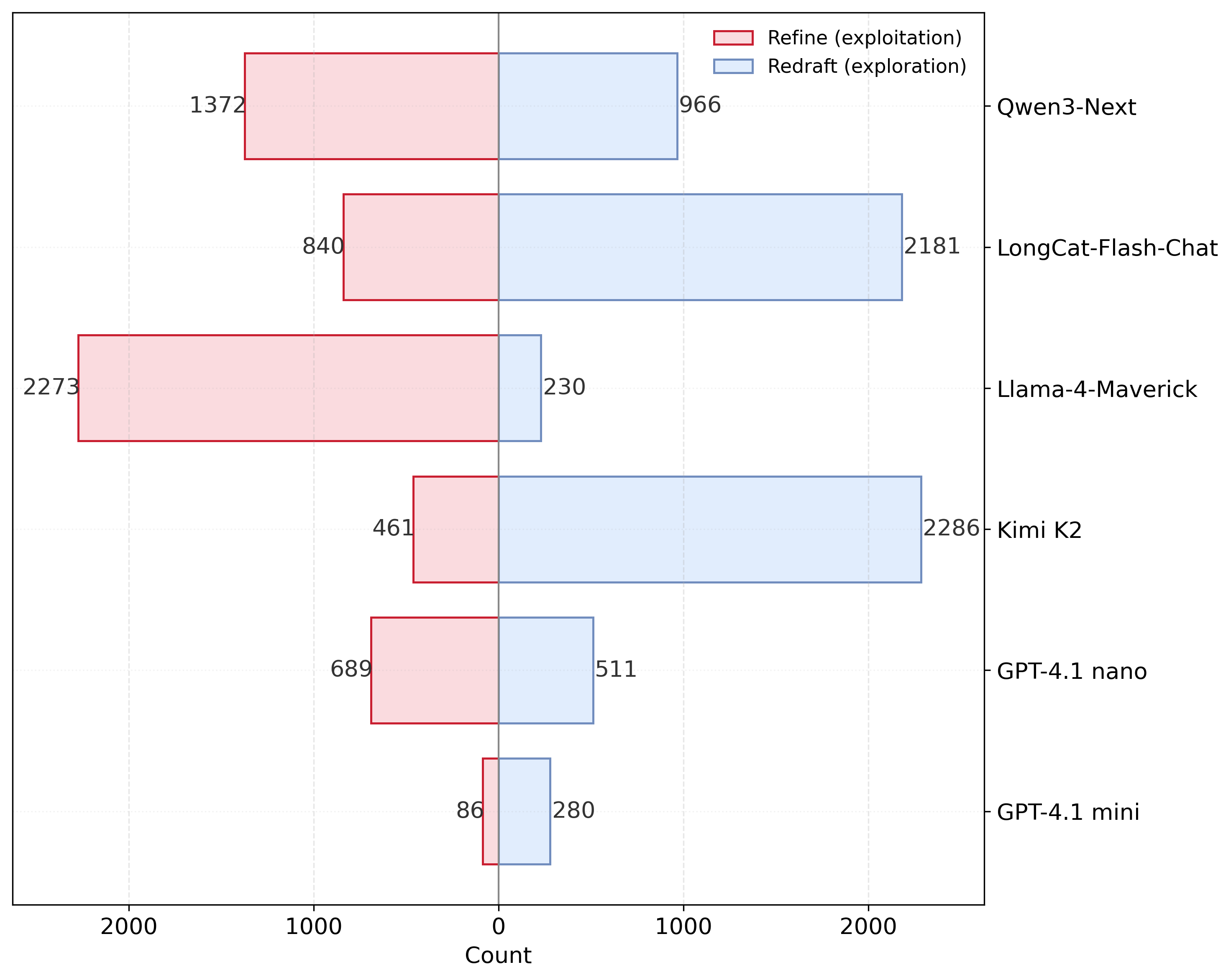}
    \caption{Butterfly bar plot depicting the count of next-action suggestions (``refine'' vs. ``redraft'') across various models within 16 iterations of the \ours framework. }
    \label{fig:butterfly}
\end{figure}
\begin{figure}
    \centering
    \includegraphics[width=\linewidth]{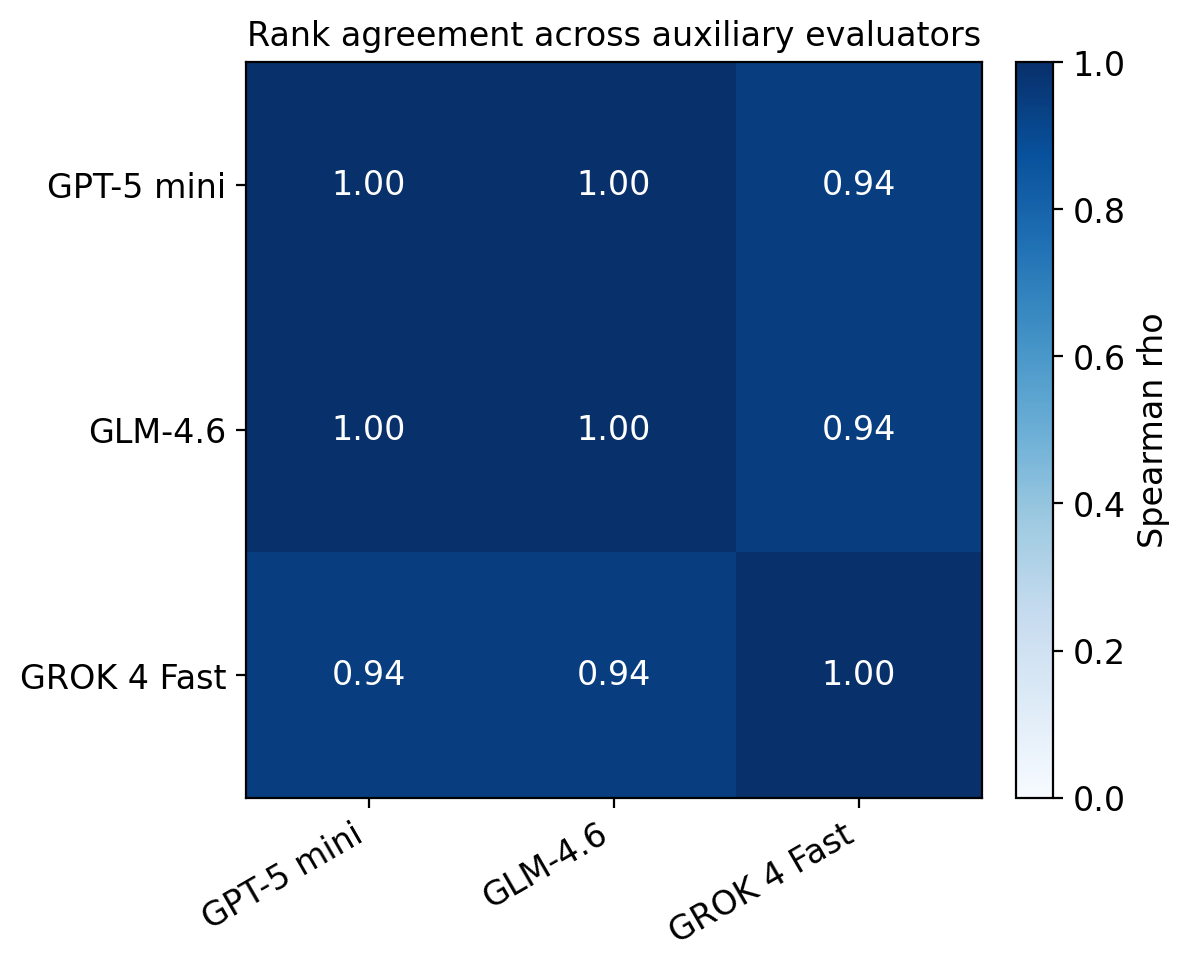}
    \caption{Ablation of ranking agreement across auxiliary evaluators. Each auxiliary model ranks generator models by Recall on Draft; the heatmap reports pairwise Spearman rank correlations ($\rho$) between these rankings, with $\rho$ shown in each cell and the diagonal fixed at 1.0. High off-diagonal $\rho$ indicates that different evaluators induce highly similar orderings over generators.}
    \label{fig:ranking}
\end{figure}

\section{Full Results}
\label{sec:full-results}
\subsection{Main Experiment}
\begin{figure*}[!htbp]
    \centering
    \includegraphics[width=\linewidth]{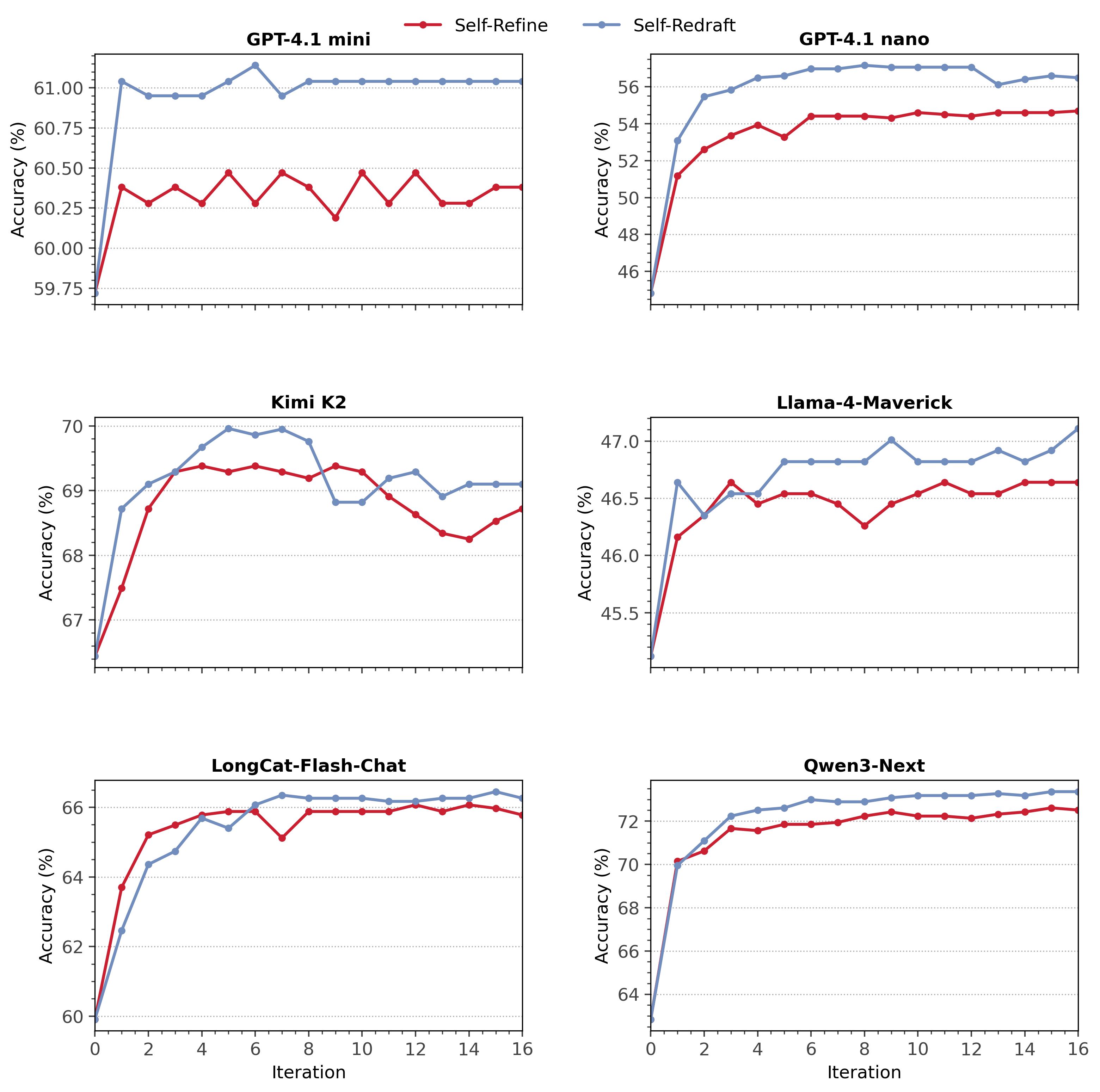}
    \caption{Iterative accuracy across six generator models: Self-Refine vs. \ours, with pass@8 as reference. }
    \label{fig:detailed_result}
\end{figure*}

Figure \ref{fig:detailed_result} demonstrates the full results of the experiment in Section \ref{subsec:main_results}.
\subsection{Blinded Evaluation}
The full result of the blinded evaluation is presented in Table \ref{tab:aux_clf}. Interestingly, we observed high consistency in the ranking of generator models by their Recall on Draft across different auxiliary models, suggesting a shared interpretation of methodological changes among the evaluators, as illustrated in Figure \ref{fig:ranking}.

\begin{table*}[!htbp]
\centering
\small
\setlength{\tabcolsep}{7pt}
\renewcommand{\arraystretch}{1.15}
\begin{tabular}{lcccccc}
\toprule
& \multicolumn{2}{c}{GPT-5 mini} & \multicolumn{2}{c}{GLM-4.6} & \multicolumn{2}{c}{GROK 4 Fast} \\
\cmidrule(lr){2-3}\cmidrule(lr){4-5}\cmidrule(lr){6-7}
Generator Model & Accuracy & Recall on Draft & Accuracy & Recall on Draft & Accuracy & Recall on Draft \\
\midrule
GPT-4.1 mini     & 56.20 & 31.39 & 57.14 & 34.13 & 55.84 & 32.85 \\
GPT-4.1 nano     & 54.45 & \textbf{51.50} & 52.97 & \textbf{55.08} & 52.30 & \textbf{50.70} \\
Kimi K2          & 56.62 & 32.10 & 57.95 & 34.85 & 55.97 & 30.59 \\
Llama-4-Maverick & 56.09 & 22.17 & 56.10 & 26.55 & 55.43 & 22.17 \\
LongCat-Flash-Chat & 55.61 & 25.10 & 56.16 & 30.56 & 57.40 & 28.46 \\
Qwen3-Next & \textbf{62.80} & 36.49 & \textbf{63.05} & 40.79 & \textbf{62.90} & 38.97 \\
\bottomrule
\end{tabular}
\caption{Blinded classification of original feedback intent (refine vs. redraft) using auxiliary models. Metrics report overall accuracy and Recall on Draft (true positive rate on redraft). Higher values indicate better recovery of the intended methodological guidance.}
\label{tab:aux_clf}
\vspace{-0.3cm}
\end{table*}

\onecolumn
\section{Case Study}
\label{sec:case study}
\begin{promptbox}[colback=black!5, colframe=white!60!black, title=Prompt Templates]{}
\textbf{\small{Task Description}}
\scriptsize
\vspace{0.2cm}

\begin{verbatim}
You are given two 0-indexed integer arrays, nums1 and nums2, both having length n.
You are allowed to perform a series of operations (possibly none).
In an operation, you select an index i in the range [0, n - 1] and swap the values of nums1[i] and nums2[i].
Your task is to find the minimum number of operations required to satisfy the following conditions:

nums1[n - 1] is equal to the maximum value among all elements of nums1, i.e., nums1[n - 1] = max(nums1[0], nums1[1], ..., nums1[n -
1]).
nums2[n - 1] is equal to the maximum value among all elements of nums2, i.e., nums2[n - 1] = max(nums2[0], nums2[1], ..., nums2[n -
1]).

Return an integer denoting the minimum number of operations needed to meet both conditions, or -1 if it is impossible to satisfy 
both conditions.

Example 1:

Input: nums1 = [1,2,7], nums2 = [4,5,3]
Output: 1
Explanation: In this example, an operation can be performed using index i = 2.
When nums1[2] and nums2[2] are swapped, nums1 becomes [1,2,3] and nums2 becomes [4,5,7].
Both conditions are now satisfied.
It can be shown that the minimum number of operations needed to be performed is 1.
So, the answer is 1.

Example 2:

Input: nums1 = [2,3,4,5,9], nums2 = [8,8,4,4,4]
Output: 2
Explanation: In this example, the following operations can be performed:
First operation using index i = 4.
When nums1[4] and nums2[4] are swapped, nums1 becomes [2,3,4,5,4], and nums2 becomes [8,8,4,4,9].
Another operation using index i = 3.
When nums1[3] and nums2[3] are swapped, nums1 becomes [2,3,4,4,4], and nums2 becomes [8,8,4,5,9].
Both conditions are now satisfied.
It can be shown that the minimum number of operations needed to be performed is 2.
So, the answer is 2.

Example 3:

Input: nums1 = [1,5,4], nums2 = [2,5,3]
Output: -1
Explanation: In this example, it is not possible to satisfy both conditions.
So, the answer is -1.


Constraints:

1 <= n == nums1.length == nums2.length <= 1000
1 <= nums1[i] <= 10^9
1 <= nums2[i] <= 10^9
\end{verbatim}

\hrule height 0.5pt
\vspace{0.3cm}

\textbf{\small{Original Solution}}
\scriptsize
\vspace{0.2cm}

\begin{verbatim}
from typing import List

class Solution:
    def minOperations(self, nums1: List[int], nums2: List[int]) -> int:
        n = len(nums1)
        max_nums1 = max(nums1)
        max_nums2 = max(nums2)
        target1 = nums1[-1]
        target2 = nums2[-1]

        # Check if it's possible to make nums1[-1] the max of nums1
        # and nums2[-1] the max of nums2 simultaneously.
        # For each position, determine if swapping is needed.
        swaps_for_nums1 = 0
        swaps_for_nums2 = 0
        total_swaps = 0

        for i in range(n):
            # If nums1[i] is less than max_nums1, to make nums1[-1] the max,
            # we need to swap at position i if swapping can help.
            if nums1[i] != max_nums1:
                # If nums2[i] is also less than max_nums1, can't bring max_nums1 to nums1[-1] here
                if nums2[i] != max_nums1:
                    # Can't achieve max_nums1 at nums1[-1]
                    return -1
                else:
                    swaps_for_nums1 += 1

            # Similarly for nums2
            if nums2[i] != max_nums2:
                if nums1[i] != max_nums2:
                    return -1
                else:
                    swaps_for_nums2 += 1

        # Now, check if the last elements are already the max
        # If not, we need to perform swaps at positions where swapping helps.
        # The minimal swaps needed is the minimum between swaps_for_nums1 and swaps_for_nums2
        # because we can choose to swap at positions that help satisfy both conditions.
        # But we need to ensure both conditions are satisfied simultaneously.
        # The minimal number of swaps is the minimum of the two counts, but only if
        # the total swaps do not exceed the total number of positions.
        # Actually, the problem reduces to choosing the minimal swaps needed to satisfy both.
        # The key insight is:
        # - For each position, if swapping helps to bring the max value to the last position,
        #   count it.
        # - The minimal swaps is the minimal number of swaps needed to fix both conditions.
        # Since swapping at position i affects both arrays simultaneously,
        # the total swaps needed is the minimum between the two counts, considering the constraints.

        # But more straightforwardly, we can count how many positions need swapping for each condition,
        # and the minimal number of swaps is the minimal between these counts, provided the total
        # swaps do not violate the possibility.

        # Actually, the problem is similar to a "minimum number of swaps to make both last elements the max"
        # which can be approached by counting positions where swapping can help.

        # Let's implement a more direct approach:
        count_swap = 0
        for i in range(n):
            # If both nums1[i] and nums2[i] are not max, but swapping can help
            if (nums1[i] != max_nums1 and nums2[i] != max_nums1) and (nums1[i] != max_nums2 and nums2[i] != max_nums2):
                # Can't fix both conditions at this position
                continue
            # Count positions where swapping helps to fix either condition
            if (nums1[i] != max_nums1 or nums2[i] != max_nums2):
                count_swap += 1

        # The minimal number of swaps needed is the minimal between fixing nums1 and nums2
        # which is the minimum of swaps_for_nums1 and swaps_for_nums2
        # but we need to ensure both conditions are satisfied simultaneously.
        # The minimal swaps needed is the minimal number of positions where swapping can fix the issues.
        # Since each swap fixes both conditions at once, the answer is the minimal number of swaps
        # needed to fix both conditions, which is the minimum between the counts.
        # But we need to check if it's possible at all.

        # Final check: if it's impossible to satisfy both conditions
        # (i.e., no positions where swapping can help), return -1
        if count_swap == 0:
            # Check if last elements are already the max
            if target1 == max_nums1 and target2 == max_nums2:
                return 0
            else:
                return -1

        # Otherwise, the minimal swaps needed is the minimum between swaps_for_nums1 and swaps_for_nums2
        # because we can choose to swap at positions that help satisfy both conditions.
        return min(swaps_for_nums1, swaps_for_nums2)
\end{verbatim}

\hrule height 0.5pt
\vspace{0.3cm}

\textbf{\small{Feedback}}
\scriptsize
\vspace{0.2cm}
\begin{verbatim}
<critique>
\end{verbatim}
The provided code attempts to solve the problem by defining a helper function `min\_swaps` that tries to count the minimal swaps needed to bring the target maximum values to the last positions of `nums1` and `nums2`. However, the logic inside `min\_swaps` is flawed and incomplete:

\begin{enumerate}[leftmargin=*]
    \item The function checks if the target values exist in either array but does not correctly handle the scenario of swapping elements at indices other than the last one to bring the targets to the end. 
    \item The code tries to count swaps by iterating over all indices except the last, but the logic for counting swaps is incorrect:
    \begin{itemize}
        \item It increments `swaps\_needed` only when both targets can be fixed at the same index, which is a very restrictive condition.
        \item It ignores cases where targets can be fixed separately at different indices.
        \item It does not consider that swapping at one index affects only that index and does not move elements closer to the end.
    \end{itemize}
    \item The approach to handle swapping at the last index by simulating a swap and calling `min\_swaps` again is not integrated well with the rest of the logic.
    \item The problem essentially requires determining if it's possible to have the maximum values at the last positions after some swaps at the same indices, and if so, the minimal number of swaps needed. The current approach does not correctly model or solve this.
    \item The problem is more about checking if the maximum values can be placed at the last index by swapping at indices where the maximum values appear, and counting how many swaps are needed to achieve this.
\end{enumerate}
In summary, the code does not correctly implement the logic to find the minimal number of swaps and will fail on the provided examples and other test cases.

A better approach would be:
\begin{itemize}[leftmargin=*]
    \item Identify the maximum values in `nums1` and `nums2`.
    \item Check if these maximum values appear at the last index or can be brought there by swapping at indices where these maximum values appear.
    \item Count the minimal number of swaps needed to place these maximum values at the last index.
    \item If impossible, return -1.
\end{itemize}
This requires careful handling of the swaps and conditions.
\begin{verbatim}
 </critique>   
\end{verbatim}
\begin{verbatim}
<suggestion>
redraft
</suggestion>
\end{verbatim}
\hrule height 0.5pt
\vspace{0.3cm}

\textbf{\small{Regenerated Solution}}
\scriptsize
\vspace{0.2cm}

\begin{verbatim}
from typing import List

class Solution:
    def minOperations(self, nums1: List[int], nums2: List[int]) -> int:
        n = len(nums1)

        # Try both possibilities for the last index:
        # 1) don't swap at last: target1 = nums1[-1], target2 = nums2[-1], cost_last = 0
        # 2) swap at last:    target1 = nums2[-1], target2 = nums1[-1], cost_last = 1
        def operations(target1: int, target2: int, cost_last: int) -> int:
            # We want after all swaps:
            # - nums1[-1] == target1
            # - nums2[-1] == target2
            # and for every i < n-1:
            #   nums1[i] <= target1 and nums2[i] <= target2
            swaps = cost_last
            for i in range(n-1):
                a, b = nums1[i], nums2[i]
                # two choices at i: no swap => (a, b), swap => (b, a)
                no_swap_ok = (a <= target1 and b <= target2)
                swap_ok    = (b <= target1 and a <= target2)
                if not no_swap_ok and not swap_ok:
                    # impossible to satisfy constraints at index i
                    return float('inf')
                # choose orientation that avoids a swap if possible
                if no_swap_ok:
                    continue
                # else we must swap
                swaps += 1
            return swaps

        # Case 1: no swap at last
        t1 = nums1[-1]
        t2 = nums2[-1]
        res1 = operations(t1, t2, 0)
        # Case 2: swap at last
        t1, t2 = nums2[-1], nums1[-1]
        res2 = operations(t1, t2, 1)

        ans = min(res1, res2)
        return -1 if ans == float('inf') else ans
\end{verbatim}
\end{promptbox}

\end{document}